\newif\ifSaveSpace
\SaveSpacefalse

\newenvironment{Proof}{\noindent{\it Proof:}\hspace*{1em}}{\qed \medskip}
\newcommand{\qed}{\rule{6pt}{6pt}}

\newcommand{\cG}{{\cal G}}
\newcommand{\cP}{{\cal P}}

\documentclass{llncs}

\usepackage{latexsym,graphics,epsfig,url}
\usepackage{amsfonts,amssymb}
\title{On Simultaneous Graph Embedding}
\author{
C.~A.~Duncan\inst{1}
\and A.~Efrat$^*$\inst{2}
\and C.~Erten$^*$\inst{2}
\and S.~G.~Kobourov\thanks{Partially supported by NSF Grant ACI-0222920.}\inst{2}
\and J.~S.~B.~Mitchell\thanks{
Partially supported by Metron Aviation, Inc., NASA Ames
Research (NAG2-1325), NSF (CCR-0098172),
and the U.S.-Israel Binational Science Foundation.}\inst{3}  }
\institute{ 
Dept. of Computer Science, Univ. of Miami, {\tt duncan@cs.miami.edu}
\and Dept. of Computer Science, Univ. of Arizona, {\tt \{alon,erten,kobourov\}@cs.arizona.edu}
\and Dept. of Applied Mathematics and Statistics, Stony Brook University, {\tt jsbm@ams.sunysb.edu}
}

\begin{document}
\date{}
\maketitle
\thispagestyle{plain}

\addtocounter{footnote}{-3}

\begin{abstract}We consider the problem of simultaneous embedding of planar graphs. There are two variants of this problem, one in which the mapping between the vertices of the two graphs is given and another where the mapping is not given. 
In particular, we show that without mapping, any number of outerplanar graphs can be embedded simultaneously on an $O(n)\times O(n)$ grid, and an outerplanar and general planar graph can be embedded simultaneously on an $O(n^2)\times O(n^3)$ grid. If the mapping is given, we show how to embed two paths on an $n \times n$ grid, or two caterpillar graphs on an $O(n^2)\times O(n^3)$ grid. 

\end{abstract}

\section{Introduction}
The areas of graph drawing and information visualization have seen significant 
growth in recent years.
Often the visualization problems %
involve taking information in the form of graphs
and displaying them in a manner that both is aesthetically pleasing and conveys some meaning.
The aesthetic criteria by itself are the topic of much debate and research,
but some generally accepted and tested standards include preferences for
straight-line edges or those with only a few bends, a limited number of crossings, 
good separation of vertices and edges, as well as a small overall area.
Some graphs change over the course of time and in such cases it is often
important to preserve the ``mental map''.
That is, slight changes in the graph structure should not yield large changes in 
the actual drawing of the graph.
Vertices should remain roughly near their previous locations and edges should be routed in 
roughly the same manner as before.

Due to the complexity of the problem with general types of graphs,
a great deal of the current theoretical research has dealt with planar graphs.
In this area, much progress has been made; see~\cite{dett-gd,mo:NishizekiChiba:88} for
an overview.
On the other hand, since the problem is NP-hard in the general case,
many of the more practical applications using general graphs have been 
studied with heuristic implementations demonstrating their effectiveness.

From a theoretical point-of-view, one recent trend in research has
been to address the topic of thickness of graphs~\cite{e-stgt-02}.  The thickness of a
graph is the minimum number of planar subgraphs into which the graph
can be partitioned.  Similar to a common technique in
VLSI design, the graph is embedded in {\em layers}.  
Any two edges drawn in the same layer can only intersect at a common
vertex, and vertices are placed are placed in the same
location across all layers.

Thus, the property of {\em graph thickness}
formalizes the notion of layered drawing of
graphs. The thickness of a graph is defined as the minimum number of
layers for which a drawing of $G$ exists, in which edges can be drawn
as Jordan curves~\cite{b-dcg-67}.
A related graph property is {\em geometric thickness}
and defined as the minimum number of layers for
which a drawing of $G$ exists, in which all edges are drawn as
straight-line segments~\cite{deh-gtcg}. Finally, {\em book thickness} of a graph
is the minimum number of layers for which a drawing
of $G$ exists, in which edges are drawn as straight-line segments and
vertices are in convex position~\cite{fk-btg-79}. 

\begin{figure}[t]
\begin{center}
\includegraphics[width=5.7cm]{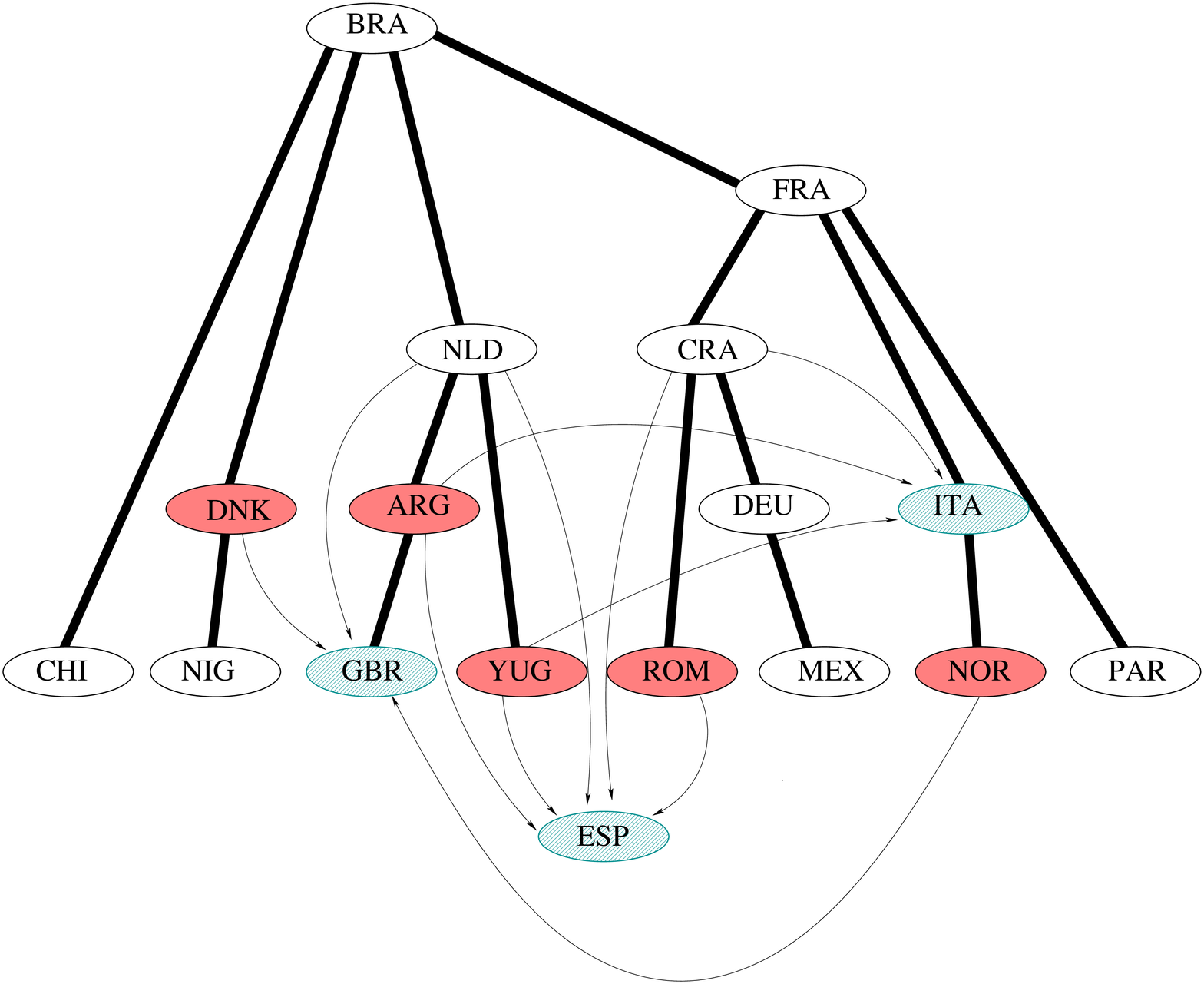}
\hspace{.4cm}
\includegraphics[width=5.7cm]{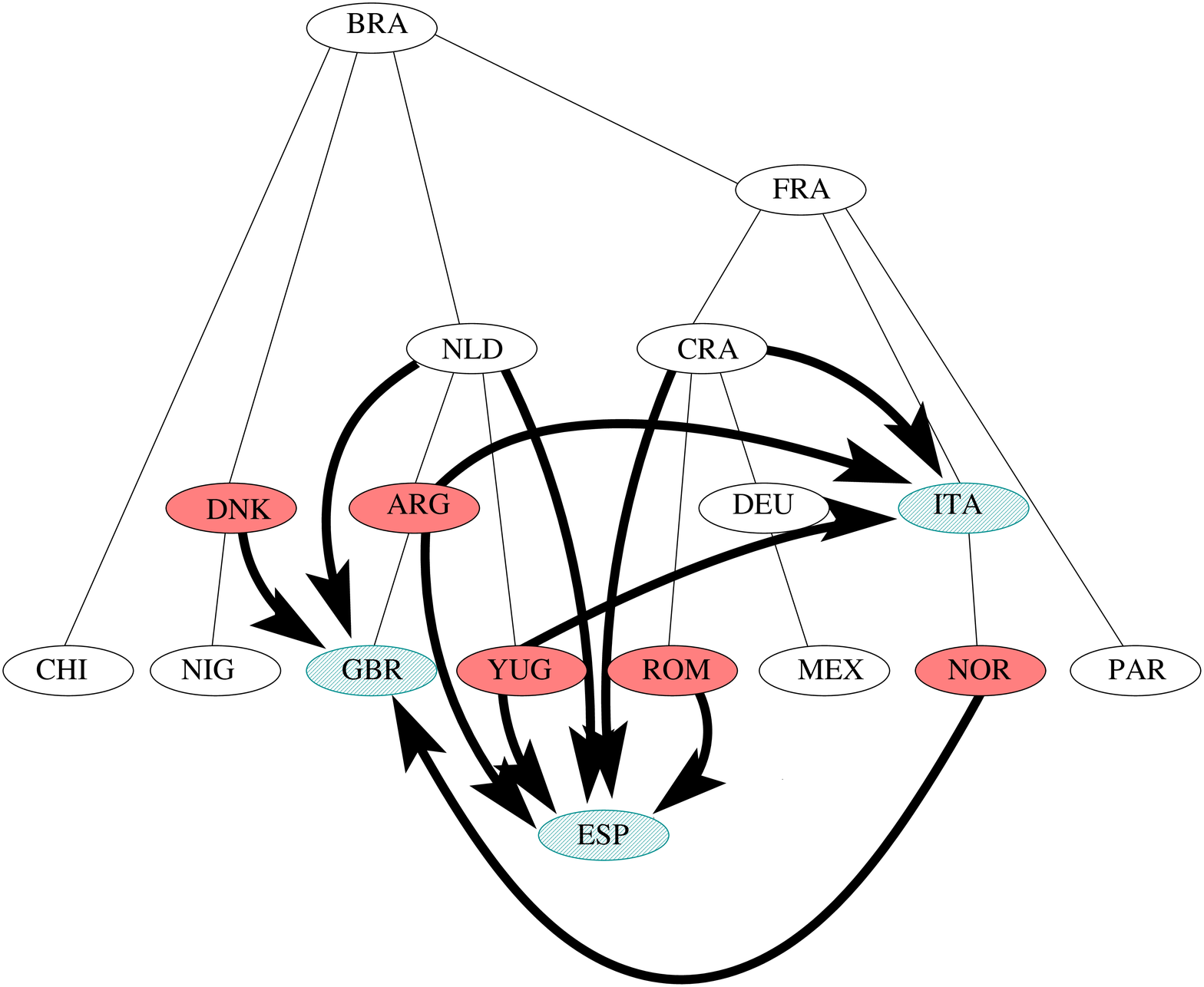}
\end{center}
\caption
{\small\sf The vertices of this graph are the round of 16 teams from Worldcup 1998 (plus Spain). The 8 teams eliminated in the round of 16 are on the bottom,
next are the 4 teams eliminated in the quarter-finals, etc.
Thick edges in the left drawing indicate matches played. Thick edges in the right drawing indicate export of
players on the club level. The light(dark) shaded vertices indicate importers(exporters) of players.}
\label{figure-soccer}
\end{figure}

We look at this problem almost in reverse.  Assume we are given the
layered subgraphs and now wish to embed the various layers so that the
vertices coincide and so that no edges cross.  We wish to
simultaneously draw the layered subgraphs, that is, we wish to display
several graphs using the same set of vertices but different sets of
edges.  Take, for example, two graphs
from the 1998 Worldcup; see Fig.~\ref{figure-soccer}. One of the graphs is a tree illustrating the
games played. The other is a graph showing the major exporters and
importers of players on the club level. In displaying the information,
one could certainly look at the two graphs separately, but then there
would be little correspondence between the two layouts if they were
created independently, since the viewer has no ``mental map'' between the
two graphs.  Using a simultaneous embedding, the vertices could be
placed in the exact same locations for both graphs and the
relationships become more clear.  This is different than simply
merging the two graphs together and displaying the information as one
large graph.

In simultaneous embeddings, we are concerned with crossings but not
between edges belonging to different layers (and thus different
graphs).  In the merged version, the graph drawing algorithm used
loses all information about the separation of the two graphs and so
must also avoid such non-essential crossings.  The techniques for
displaying simultaneous embeddings are also quite varied.  One may
choose to draw all graphs simultaneously, employing different edge
styles, colors, and thickness for each edge set.  One may choose a
more three-dimensional approach to giving a visual difference between
layers.  One may also choose to show only one graph at a time and
allow the users to choose which graph they wish to see; however, the
vertices would not move, only the edge sets would change.  Finally, one
may highlight one set of edges over another giving the effect of
bolding certain graphs.

The subject of simultaneous embeddings has many different variants,
several of which we address here.  The two main classifications we
consider are embeddings with (and without) predefined vertex mappings.
\begin{itemize}
\item {\bf Simultaneous (Geometric) Embeddings With Mapping:} 
Given $k$ planar graphs $G_i(V, E_i)$ for $1 \leq i \leq k$, find plane
straight-line drawings $D_i$ of $G_i$ such that for every 
$v\in V$ and any two drawings $D_i$ and $D_j$, $v$ is mapped to the same
point on the plane in both drawings.
\footnote{Note that the vertices in each graph are uniquely mapped so that they may
be treated as the same vertex set.}
\item {\bf Simultaneous (Geometric) Embeddings Without Mapping:} 
Given $k$ planar graphs $G_i(V_i, E_i)$ for $1 \leq i \leq k$, find 
a one-to-one mapping among all pairs of vertices in $V_i, V_j$ such that
a simultaneous (geometric) embedding with this mapping exists.
\end{itemize}

The problem of simultaneous embedding of $k$ graphs with given mapping
can be seen as testing whether the combined graph (in which the edges
set is the union of the edge sets of the given graphs) has geometric
thickness $k$. That is, if the $k$ graphs can be embedded
simultaneously, then the thickness of the combined graph is at most $k$,
while the opposite direction is not necessarily true. A similar
relationship exists between the problem of simultaneous embedding of a
graph when the mapping is not given and testing for the geometric
thickness of the combined graph.

Note that in the final drawing a crossing between two edges $a$ and
$b$ is allowed {\em only if} there does not exist an edge set $E_i$
such that $a,b \in E_i$.  The following table summarizes our current
results regarding the two versions under various constraints on the
type of graphs given. Each column indicates the size of the integer
grid required, with the expression ``not possible'' indicating that
there exist graphs of this category that do not have embeddings.

\vspace{.2cm}
{
\scriptsize
\begin{center}
\begin{tabular}{||l|c|c|c|c||}
\hline
	{\bf Graphs}		& {\bf With Mapping} 	& {\bf Without Mapping}\\
\hline
\hline $G_1, G_2$: Planar 	& not possible		& ? \\
\hline $G_1, G_2$: Outerplanar	& ?			& $O(n) \times O(n)$\\
\hline $G_1$: Planar, $G_2$: Outerplanar
				& ?			& $O(n^2) \times O(n^3)$\\
\hline $C_1, C_2$: Caterpillar
				& $O(n^2) \times O(n^3)$	& N/A\\
\hline $C_1$: Caterpillar, $P_2$: Path
				& $O(n) \times O(n)$	& N/A\\
\hline $P_1, P_2$: Path		& $n \times n$		& N/A\\
\hline $P_1, P_2, \dots, P_5$: Path
				& not possible		& N/A\\
\hline $T_1, T_2, T_3$: Tree	& not possible
& N/A\\
\hline\hline
\end{tabular}
\end{center}
}

\section{Previous Work}
Computing straight-line embeddings of planar graphs on the integer grid is a well-studied
graph drawing problem. The first solution to this problem is given by
de Fraysseix, Pach and Pollack~\cite{fpp-sssfe-88} using a canonical 
labeling of the vertices in an algorithm that embeds a planar graph 
on $n$ vertices on the $(2n-4)\times (n-2)$ integer grid. 
Schnyder~\cite{s-epgg-90} presents a barycentric coordinates 
method that reduces the grid size to $(n-2)\times (n-2)$. The 
algorithm of Chrobak and Kant~\cite{chrobak97convex} embeds a 
3-connected planar graph on a $(n-2)\times (n-2)$ grid so that 
each face is convex. Miura, Nakano, and Nishizeki~\cite{nakano4connected01} 
further restrict the graphs under consideration to 4-connected 
planar graphs and present an algorithm for straight-line embeddings 
of such graphs on a $(\lceil n/2\rceil-1)\times (\lfloor n/2\rfloor)$ grid.

Another related problem is that of simultaneously embedding more than
one planar graph, not necessarily on the same point set. 
In a paper dating back to 1963, Tutte~\cite{t-hdg-63} shows that 
there exists a simultaneous straight-line representation of a planar 
graph and its dual in which the only intersections are between 
corresponding primal-dual edges. Bern and Gilbert~\cite{Bern:1992:DPD} 
address a variation of the problem: finding suitable locations for dual
vertices, given a straight-line planar embedding of a planar graph, so
that the edges of the dual graph are also straight-line segments and cross
only their corresponding primal edges. They present a linear time
algorithm for the problem in the case of convex 4-sided faces and
show that the problem is NP-hard for the case of convex 5-sided
faces. Erten and Kobourov~\cite{ek-sepg-02} present an algorithm 
for embedding a planar graph and its dual on the $O(n)\times O(n)$ grid so 
that both graphs are drawn with straight lines, every dual vertex is 
inside its corresponding face, and the only crossings are between 
primal-dual edge pairs.

\section{Simultaneous Embedding With Mapping}

\begin{definition}
Let $G_i(V, E_i)$ for $i \in \{1, \dots, k\}$ be $k$ graphs with the 
same vertex set but different edge sets.
A $k$-layer geometric embedding of $\cG = \bigcup G_i$ is a straight 
line drawing of $\cG$ such that two edges intersect only at a common vertex or if they belong 
to different edge sets.
\end{definition}

This broad description of the problem has many different variants and 
many questions still unanswered. We first address the simplest 
problem: embedding paths.

\begin{theorem}
Let $P_1$ and $P_2$ be 2 paths on the same vertex set of size $n$.
Then a $2$-layer geometric embedding of $P_1$ and $P_2$ can be found in linear time and 
on an $n \times n$ grid.
\label{the-two-path-theorem}
\end{theorem}

\begin{figure}[t]
\begin{center}
\ifSaveSpace
	\centerline{Insert Figure: 2-paths}
\else
	\scalebox{.8}{\input{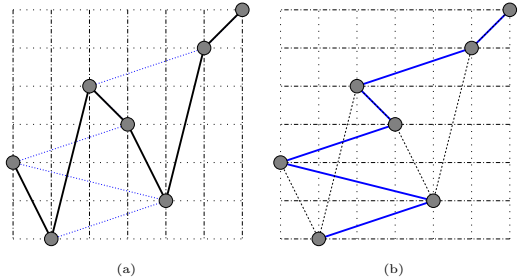}}
\fi
\end{center}
\caption
{\small\sf An example of embedding two paths on an $n \times n$ grid.
The two paths are respectively $v_1, v_2, v_3, v_4, v_5, v_6, v_7$ and
$v_2, v_5, v_1, v_4, v_3, v_6, v_7$.
They are drawn using (a) increasing $x$-order and (b) increasing $y$-order.
Notice that in (a) the vertices can slide up and down the grid's vertical ``bars''
without introducing any crossings.
A similar feature happens in (b).}
\label{fig:two-paths}
\end{figure}

\begin{Proof}
Observe that if a path is placed with neighboring vertices in increasing $x$-order, then 
the embedding will always be a straight-line drawing with no crossings 
regardless of the value of $y$. That is, if $v$ and $w$ are two vertices 
on path $p$ such that $v$ comes before $w$ in the path then $v(x) < w(x)$.
The same relationship holds if we interchange $x$ and $y$; see Fig.~\ref{fig:two-paths}.
For a vertex $v$ let $p_i$ be the vertex's position in the path 
$P_i$, $i \in \{1,2\}$. Then the vertex $v$ is placed at grid 
position $(p_1, p_2)$.
Thus, vertices in the first path are positioned in increasing $x$-order,
and vertices in the second path are positioned in increasing $y$-order.
From the above observation, we know that neither path is self-intersecting.
\end{Proof}

This simple algorithm does not seem to extend to more than two paths. 
Moreover, we can prove that five paths cannot be simultaneously embedded 
(using a given vertex mapping).

\begin{theorem}
There exist five paths $\cP = \bigcup_{1\leq i \leq 5} P_i$ on the same 
vertex set $V$ such that at least one of the layers {\em must} have a crossing.
\end{theorem}

\begin{Proof}
A path of $n$ vertices is simply on ordered sequence of $n$ numbers.
This makes identifying paths a bit easier.
The five paths are quite simple: 12345, 13542, 25134, 32415, 35214.
For example, the sequence $13542$ represents the path $(v_1, v_3, v_5, v_4, v_2)$.

We first point out that any five-vertex path is 
a subset of the complete graph of 5 vertices, $K_5$.
In fact, the union of the five five-vertex paths is a subset of $K_5$.
Thus, any $5$-layer embedding will be a subset of some drawing of $K_5$.
Since $K_5$ cannot be drawn without a crossing, at least two edges, say $a$ and $b$, must cross.
This does not immediately imply that a path crosses itself,
only that two paths could cross each other.
Since the paths are subsets of $K_5$ they would only self-intersect if edges 
$a$ and $b$ were both present in the same path.

However, the choice of edges $a$ and $b$ is up to the embedder.
Therefore, we must construct a set of paths such that any pairing of edges 
$a$ and $b$ occurs in at least one path.
For example, let us assume that $a=(v_1,v_2)$
and $b=(v_4,v_5)$, and that the five vertices are placed such 
that $a$ and $b$ cross. For notation, this pairing is labeled as $12-45$.
Then, we must guarantee that there is at least one path in our set such that 
the sequence $12$ (or $21$) and $45$ (or $54$) both appear.
Since two edges cannot intersect if they share a vertex in common,
all four vertices in the pairing must be distinct, 
So, there are 15 possible edge pairings for $K_5$.
Any path of length five can ``eliminate'' at most three pairings.
For example, the path $12345$ eliminates three pairings, $12-34$, $12-45$, 
and $23-45$, and no others. Thus, we need a {\em minimum} of five paths to 
guarantee a crossing.

A careful study of the five paths above reveals that all such edge 
pairings are present in at least one path. Therefore, these five paths cannot 
simultaneously be embedded without at least one path self-crossing.
\end{Proof}

A simple consequence of this proof is that any four paths of five vertices 
{\em can} be embedded. Thus, if were to show that three or four paths can not always
be simultaneously embedded we must use paths with more than five vertices.
We, therefore, must leave the question 
open for three and four paths. We now look at a few more variants of the problem.

\subsection{Caterpillars}

A caterpillar graph is a simple tree consisting of a path of vertices and a set of zero or more legs.  
Let us first define the specific notion of a caterpillar graph.

\begin{figure}[t]
\begin{center}
\includegraphics[width=6cm]{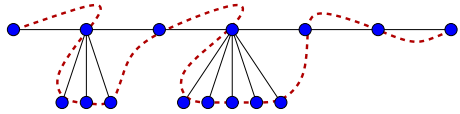}
\end{center}
\caption{\small\sf 
A caterpillar graph $C$ is drawn with solid edges. The vertices on the top row form the spine and the vertices on the bottom row form the legs of the caterpillar. The caterpillar can be traversed  in the order indicated by the dashed curve, thus creating the path $P(C)$.
}
\label{figure-caterpillar}
\end{figure}
\begin{definition}
A caterpillar graph $C=(V,E)$ is a tree consisting of two sets of vertices $V_p = \{p_1, p_2, \dots, 
p_{n'}\}$ and $V_l = \{l_1, l_2, \dots, l_{n-n'}\}$ and two sets of edges 
$E_p = \{p_1p_2, p_2p_3, \dots, p_{n'-1}p_{n'}\}$ and $E_l=\bigcup_{l \in V_l} \{p_p(l)l\}$ 
where $p(l)$ is the (unique) parent of $l$ in $V_p$.
We denote $V_p$ as the set of {\em parent} vertices and $V_l$ as the set of leg vertices.
Furthermore, for notation we let $l_{ij} \in V_l$ be the $i$-th vertex with parent $p(l_{ij}) = p_j$.
We shall often refer to the path formed by $(V_p, E_p)$ as the {\bf spine} of $C$; see Fig.~\ref{figure-caterpillar}.
\end{definition}

\begin{lemma}
\label{lemma-general-position}
Let $S=\{p_1, p_2, \dots, p_n\}$ be a set of $n$ points located on an $O(n)\times O(n)$ integer grid. 
We can refine the grid into an $O(n^2)\times O(n^3)$ grid  so that no three points are collinear. 
\end{lemma}

\begin{Proof}
We start by making a square $S_i$ of unit side length, centered around each point $p_i$. Now
we decompose each square $S_i$ into an $O(n)\times O(n^2)$ grid, so that the overall grid size is 
$O(n^2)\times O(n^3)$.   
We will now proceed to relocate each point $p_i$ inside $S_i$ incrementally so that $p_i$ is not collinear
with any 
two (previously placed) points $p_j, p_k$ with $j,k < i$.

We can do this for the first two points, leaving them in their original positions.
Thus, let us assume we have placed the first $i-1$ points and wish to place $p_i$.
There are exactly ${i-1 \choose 2} < i^2$ lines defined by the previously placed points.
Let $h$ be the number of horizontal lines
passing through $S(i)$. Each horizontal line through $S(i)$ kills 
$O(n^2)$ grid points and every other line kills $O(n)$. Then collectively $O(i^2)$ pairs 
kill $h\times n^2+(i^2-h)\times n$ points. Since $h$ is at most $i/2$, in total
$O(n^3)$ grid points of $S(i)$ get killed which leaves us with an option to displace
$p_i$ so that it is not collinear with two other displaced vertices, and by induction
we are done.
\end{Proof}

\begin{theorem}
Any two caterpillar graphs have a 2-layer geometric embedding with grid size $O(n^2) \times O(n^3)$.
\end{theorem}

\begin{Proof}
We first transform each caterpillar graph $C_1$ and $C_2$ into a path.
Let $C = (V_p\bigcup V_l, E_p \bigcup E_l)$ be a caterpillar and let $d(i)$ represent the "degree" of
$p_i \in V_p$ as the number of children in $V_l$ with parent $p_i$..
We connect the path with the following vertex ordering $p_1, l_{11}, l_{12}, \dots, l_{1d(1)}, 
p_2, l_{21}, \dots, l_{2d(2)},  \dots, p_{n'}, l_{n'd(n')}$.
That is, we start with the first parent, then its children in order, then the next parent, and its children,
until
we have the entire path constructed.
Let $P(C)$ denote the converted path graph.

Let $P(C_1)=P_1$ and $P(C_2)=P_2$ denote the two paths formed from the caterpillar graphs.
We use the technique from Theorem~\ref{the-two-path-theorem} to embed the two paths on
an $n \times n$ grid.
We now increase the grid resolution to $O(n^2) \times O(n^3)$ by taking each unit grid location 
in the original
embedding 
and refining it into an $O(n) \times O(n^2)$ grid.
Notice that the paths are still embedded properly as the only condition is that each point in a path be in 
increasing $x$ (or $y$) order.

We now apply Lemma~\ref{lemma-general-position} to yield a new embedding of the paths such that the points
are 
guaranteed to be in general position.
It is easy to see that the paths are still properly placed 
since each point is placed inside a square of unit length centered around
the original point location so that the relative $x$ (and $y$)order of the points remain the same.

Now, we construct the caterpillar embedding by using the same points and removing all edges replacing them
with the 
edges from the caterpillar.
To see that the graph is properly embedded we simply need to look at any parent vertex $p_i$ and prove none
of its 
edges intersect any other edges.
This is where the ordering of the path becomes significant.
Without loss of generality, let us look at the caterpillar $C$ whose path was formed along the 
$x$ direction.
Since no three points are collinear, we can connect all "leg" edges between $p_i$ and its children
as well as 
connecting $p_i$ with $p_{i+1}$ (and symmetrically $p_{i-1}$ without introducing
any crossing among these edges.
Let $p_j$ be any parent vertex such that $j < i$; i.e, $p_j$ came before $p_i$.
From the embedding technique, we know that the x-coordinate of $p_j$ is less than $p_i$'s.
If we look at the edge from $p_j$ to $p_{j+1}$ either $j+1 = i$ or else the edge from $p_j$ 
to $p_{j+1}$ must
lie 
complete to the left of $p_i$ and its edges and thus not intersect.
Similarly, if we look at the edge from $p_j$ to any of its children, which also come before 
$p_i$ in the path
ordering, 
we see that that edge must also lie completely to the left of $p_i$ and its edges.
Therefore, the caterpillar graph is properly embedded and we are done.
\end{Proof}

We believe that this space constraint can be reduced. 
As a first step in this direction we argue that a path and a caterpillar can be embedded in a much smaller area as the following theorem shows.

\begin{theorem}
Given a path $P$ and a caterpillar graph $C$, we can simultaneously embed them on an $n \times 2n$ grid.
\end{theorem}

\begin{Proof}
We embed these using much the same method as embedding two paths with one exception, that we allow
some vertices to share the same $x$-coordinate.
For a vertex $v$ let $o_p(v)$ denote $v$'s position in $P$.
If $v$ is in $V_p$, i.e. a parent vertex in $C$, then let $o_c(v)$ be its position among the parent path,
disregarding the legs,
and $v$ is initially embedded at the location $(2o_c(v), o_p(v))$.
Otherwise, $v \in V_l$.
Let $o_c(v) = o_c(p(v))$ be its parent's position and initially embed $v$ at the location $(2o_c(v)+1,
o_p(v))$.

We now proceed to attach the edges.
Clearly the path works properly if we preserve the $y$-ordering of the points.
In our embedding, we may need to shift, but we shall only perform right shifts.
That is, we shall push points to the right of a vertex $v$ by one unit right, in essence inserting one extra
grid location when necessary.
Notice this step will preserve the $y$-ordering.
To attach the caterpillar edges, let us march along the spine.
We shall guarantee that for any parent vertex $p_i$ that no edges extending from a vertex $p_j$ with $j < i$
intersect with any
other placed vertices.
That is, none of the edges up to $p_i$ intersect.
Clearly, this works for $p_1$ as there are no edges yet placed.
Let us then assume it holds for $p_i$.
To show it hold for $p_{i+1}$ we must properly place $p_i$ so that none of its edges intersect any others.
Since all of the legs of $p_i$ lie on the same $x$-coordinate but with differing $y$-coordinates 
and one unit to the right of $p_i$, we can connect $p_i$ to its legs without any crossings.
Also notice that all other previously placed edges lied to the left of $p_i$ and again there could be no
crossings.
We must now connect $p_i$ to $p_{i+1}$.
However, it is possible that at most one leg is collinear with $p_i$ and $p_{i+1}$.
In this case, we simply shift $p_{i+1}$ and all succeeding points by one unit to the right.
We continue the right shift until none of the legs is collinear with $p_i$ and $p_{i+1}$.
Now, the edge $p_i$ to $p_{i+1}$ does not intersect another vertex. The number of shifts we made is bounded
by $k_i$, where $k_i$ is the number of legs of parent vertex $p_i$.  

We continue in this manner until we have attached all edges.
Let $k$ be the total number of legs of the caterpillar. Then the total number of shifts 
made is $k$. Since we initially start with $2\times (n-k)$ columns in our grid, the total number
of columns necessary is $2n-k$. Thus, in the worst case the grid size is less than $2n \times n$.
\end{Proof}

\begin{theorem}
There exists a set of three caterpillar graphs that cannot be simultaneously embedded.
\end{theorem}

\begin{Proof}
The result is a direct interpretation of Theorem 2 from~\cite{e-stgt-02} 
along with its proof.
The author proves that there exists a graph of (graph) thickness three 
with geometric thickness greater than three.
The graph, defined as $G_3(n)$ is defined on $n + {n \choose 3}$ vertices representing
all possible singleton and tripleton subsets of an $n$-element set.
Edges are formed between each tripleton vertex and the three singleton vertices 
forming the subset. Since this is a bipartite graph and the tripleton vertices have 
degree exactly three, we
can partition this graph into three subgraphs such that each subgraph is a forest of 
stars, the centers
being the singleton vertices and the three edges from a tripleton vertex belonging to 
different subgraphs.
The authors then prove that there exist values of $n$ such that $G_3(n)$ has geometric 
thickness
greater than three.
This implies that the three subgraphs of stars cannot be simultaneously embedded using 
straight-line edges without crossings.
In order that each subgraph be a tree instead of a forest,
we can readily add edges to connect the components while not changing the
graph thickness or the geometric thickness.
Therefore, the three 
caterpillar graphs do not have geometric thickness three.
\end{Proof}

The theorem, although disappointing, still leaves open the question of whether two general 
trees can be simultaneously embedded even if they have bounded (not necessarily constant) degree.

\begin{figure}[t]
\begin{center}
\includegraphics[width=8cm]{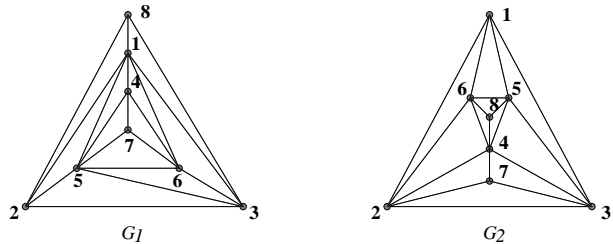}
\end{center}
\vspace{-.5cm}
\caption{\sf Given the above mapping between the vertices; $G_1$ and $G_2$
can not be embedded simultaneously. If a simultaneous embedding were possible
that embedding would force three vertices on the outer face. However 
the above graphs do not share any faces.}
\label{general}
\end{figure}

\subsection{General Planar Graphs}

Simultaneous embedding of general planar graphs is not always possible.

\begin{theorem}
There exist two general planar graphs which, given a mapping between the vertices 
of the graphs, cannot be simultaneously embedded.
\end{theorem}

\begin{Proof}
As Figure~\ref{general} illustrates, regardless of the embedding chosen there must
exist at least one external triangular face shared by both, but the graphs do not
have any faces in common.
\end{Proof}


\section{Simultaneous Embedding Without Mapping}
In the following subsections we present methods to embed different classes of graphs
simultaneously when no mapping between the vertices are provided. For the remainder of 
this section, when we say simultaneous embeddings we always mean without vertex
mappings.

\subsection{A General Planar Graph and an Outerplanar Graph}
The following theorem summarizes our results on simultaneously embedding 
an outerplanar graph and a general planar graph.
\begin{theorem} 
\label{general-outerplanar}
Two planar graphs $G_1$ and $G_2$ each with $n$ vertices can be
simultaneously embedded (without mapping) on an $O(n^2)\times O(n^3)$
grid if one of the graphs is outerplanar.
\end{theorem}

\subsubsection{General Position Requirement}
We start by showing how to find a straight-line embedding 
of a planar graph $G$ on the 
grid with the {\it{general position requirement}}, i.e., no 
three vertices of $G$ lie on a line. We first present an algorithm
that requires $O(n^3)\times O(n^4)$ size grid. Then we improve the grid size to be
$O(n^2)\times O(n^3)$. 

We begin by drawing the given graph $G$ in an $O(n)\times O(n)$
integer grid, $I$. This step can be done using one of the 
grid-drawing algorithms~\cite{chrobak97convex,fpp-sssfe-88,s-epgg-90}. 
The resulting drawing has potentially many collinearities among the 
vertices. However, any vertex $w$ that is {\em not} collinear with a
line through vertices $u$ and $v$ must be at least $1/n\sqrt{2}$ away 
from the line $uv$. We leave the details of this claim out of the abstract.

Next, we center an axis-aligned square, $S(v_i)$, of side length $1/n\sqrt{2}$
on each vertex $v_i$.  By the property above, if vertices $u$, $v$, and
$w$ are {\em not} collinear in the embedding of $G$, then $u'$, $v'$,
and $w'$ are also not collinear, for any choice of $u'\in S(u)$, $v'\in
S(v)$, and $w'\in S(w)$.  Now, similar to Lemma~\ref{lemma-general-position}, 
we decompose each square $S(v_i)$ into an
$O(n)\times O(n^2)$ grid, so that each $v_i$ can be
displaced within the fine grid of the small neighborhood $S(v_i)$,
allowing us to {\it break} any existing collinearities.  By our choice of
the small neighborhoods, we know that we do not {\it create} any new
collinearities among vertices in the embedding.  Thus, an
$O(n^3)\times O(n^4)$ grid is sufficient for a general position
embedding of a planar graph on $n$ vertices.

\subsubsection{Improving the Grid Size}
In order to improve the grid size we make the following observation: 
In the preceding argument the necessary condition was  
that given $u$, $v$, $w$ that are not collinear, for any choice of
$u'\in S(u)$, $v'\in S(v)$, and $w'\in S(w)$ then $u'$, $v'$,
and $w'$ are also not collinear. However we can relax this condition
by insisting that it holds only if there exists an edge $(u, v)$, $(v, w)$
or $(u, w)$. If there is no such edge then picking any point 
in $S(u)$, $S(v)$, and $S(w)$ does not change the embedding of the graph. 

We use the algorithm of~\cite{cgt-cdgtt-96} to initially draw the
graph to guarantee this new property. The algorithm draws every
3-connected planar graph in an $O(n)\times O(n)$ grid under the {\it
edge resolution rule}, which guarantees that the minimum distance
between an edge and a non-incident edge or vertex is at least one grid unit.

Given this drawing, we center an axis-aligned square, $S(v_i)$, of unit side 
length on each vertex $v_i$. By the edge resolution rule, if there exists an edge
$(u, v)$ and vertices $u$, $v$, and
$w$ are {\em not} collinear in the embedding of $G$, then $u'$, $v'$,
and $w'$ are also not collinear, for any choice of $u'\in S(u)$, $v'\in
S(v)$, and $w'\in S(w)$. We now decompose each square $S(w)$ into an
$O(n)\times O(n^2)$ grid as before and the rest of the argument follows 
as before, yielding the desired result.

\begin{lemma} 
\label{graph-on-general-position}
A planar graph on $n$ vertices can be embedded in 
an $O(n^2)\times O(n^3)$ grid so that no three vertices are
collinear.
\end{lemma}

\subsubsection{Embedding Outerplanar Graphs}
Given a graph $H=(V,E)$ with $k$ vertices, and a point set $P$ with $k$ points 
on the plane, we say that $H$ can be {\it{straight-line embedded}} onto $P$, 
if there exists a one-to-one mapping $\phi :V\rightarrow P$, from the 
vertices of $H$ onto the points of $P$ such that edges of $H$ intersect 
only at vertices. The largest class of graphs known to admit such a 
straight-line embedding is the class of outerplanar graphs. Gritzmann 
et al~\cite{Gritzmann91} provide an embedding algorithm for such graphs 
that runs in $O(k^2)$ time. Bose~\cite{Bose97} further reduces the running time to 
$O(k\lg^{3}{k})$.
\begin{lemma}~\cite{Bose97,Gritzmann91} 
\label{outerplanar}
Given an arbitrary set $P$ of $k$ 
points in the plane, no three of which lie on a line, an outerplanar graph 
$H$ with $k$ vertices can be straight-line embedded on $P$. 
\end{lemma}

\subsubsection{Simultaneous Embedding}
Lemma~\ref{graph-on-general-position} and Lemma~\ref{outerplanar} together provide an 
algorithm for simultaneously 
embedding planar graphs and Theorem~\ref{general-outerplanar} follows. 
If a mapping is not given, two or more outerplanar graphs can be simultaneously embedded. This result is summarized in the theorem below.

\begin{theorem}
Any number of outerplanar graphs can be simultaneously embedded (without mapping) on an $O(n) \times O(n)$ grid.
\end{theorem}

\begin{Proof}
Since outerplanar graphs can be embedded on a pointset with points in general position (\ref{outerplanar}) it suffices to show that we can find $n$ points in general position from a grid of size $O(n) \times O(n)$. 
Take any prime number $p$ slightly larger than $n$
then take the points $(t, t^2 \mbox{ mod } p)$ for $t=1,\ldots,p$.
These points are in general position.
\end{Proof}

\section{Open Problems}
The following extensions of this work remains open:

\begin{itemize}
\item Can general planar graphs be simultaneously embedded without mapping?
\item Can 3 or 4 paths be embedded simultaneous with mapping?

\end{itemize}

\section{Acknowledgments}

We would like to thank Anna Lubiw for introducing us to the problem of
simultaneous graph embedding. We would also like to thank Ed
Scheinerman, Carola Wenk, Peter Brass, and Dean Starrett for stimulating discussions about
different variations of the problem.

\bibliographystyle{abbrv}
{\small
\bibliography{stephen.bib}
}
\end{document}